\documentclass[12pt,preprint]{aastex}
                                                         
\begin{document}                        

\title{Direct Detection of the Close Companion of Polaris with the {\it
Hubble Space Telescope}\altaffilmark{1}}                     
      
\author{Nancy Remage Evans,\altaffilmark{2}
Gail H. Schaefer,\altaffilmark{3} 
Howard E. Bond,\altaffilmark{3}
Giuseppe Bono,\altaffilmark{4}
Margarita Karovska,\altaffilmark{2}  
Edmund Nelan,\altaffilmark{3}
Dimitar Sasselov,\altaffilmark{5}
and   
Brian D. Mason\altaffilmark{6} 
}  

\altaffiltext{1}
{Based on observations with the NASA/ESA {\it Hubble Space Telescope\/} obtained
at the Space Telescope Science Institute, which is operated by the Association
of Universities for Research in Astronomy, Inc., under NASA contract
NAS5-26555.}

\altaffiltext{2}
{Smithsonian Astrophysical Observatory,    
60 Garden St., Cambridge, MA 02138; nevans@cfa.harvard.edu}

\altaffiltext{3}
{Space Telescope Science Institute, 3700 San Martin Drive, Baltimore, MD
21218; schaefer@chara-array.org, bond@stsci.edu, nelan@stsci.edu}

\altaffiltext{4}
{INAF--Osservatorio Astronomico di Roma, Via Frascati 33, 
00040, Monte Porzio Catone, Italy;
European Southern Observatory, Karl-Schwarzschild-Str.~2,
85748 Garching bei M\"unchen, Germany; bono@mporzio.astro.it}

\altaffiltext{5}
{Harvard University,    
60 Garden St., Cambridge, MA 02138}

\altaffiltext{6}
{U.S. Naval Observatory, 3450 Massachusetts Avenue, NW, Washington, DC, 20392; 
bdm@usno.navy.mil}

\begin{abstract}

Polaris, the nearest and brightest classical Cepheid, is a single-lined
spectroscopic binary with an orbital period of 30 years. Using the High
Resolution Channel of the Advanced Camera for Surveys onboard the {\it Hubble
Space Telescope\/} ({\it HST\/}) at a wavelength of $\sim$2255~\AA, we have
directly detected the faint companion at a separation of $0\farcs17$. A second
{\it HST\/} observation 1.04~yr later confirms orbital motion in a retrograde
direction. By combining our two measures with the spectroscopic orbit of Kamper
and an analysis of the {\it Hipparcos\/} and FK5 proper motions by Wielen et
al., we find a mass for Polaris~Aa of $4.5^{+2.2}_{-1.4} \, M_\odot$---the 
first purely dynamical mass determined for any Cepheid. For the faint companion
Polaris~Ab we find a dynamical mass of $1.26^{+0.14}_{-0.07} \, M_\odot$,
consistent with an inferred spectral type of F6~V and with the flux difference
of 5.4~mag observed at 2255~\AA\null. The magnitude difference at the $V$ band
is estimated to be 7.2~mag. Continued {\it HST\/} observations will
significantly reduce the mass errors, which are presently still too large to
provide critical constraints on the roles of convective overshoot, mass loss,
rotation, and opacities in the evolution of intermediate-mass stars. 

Our astrometry, combined with two centuries of archival measurements, also
confirms that the well-known, more distant ($18''$) visual companion, Polaris~B,
has a nearly common proper motion with that of the Aa,Ab pair. This is
consistent with orbital motion in a long-period bound system. The ultraviolet
brightness of Polaris~B is in accordance with its known F3~V spectral type if it
has the same distance as Polaris~Aa,Ab.

\end{abstract}

\keywords{astrometry --- binaries: visual --- Cepheids --- stars: fundamental
parameters --- stars: individual (Polaris) }

\section{Introduction}

Cepheid variable stars are of central importance in galactic and extragalactic
astronomy.  They are the primary standard candles for measuring extragalactic
distances, and they provide critical tests of stellar-evolution theory. 
Surprisingly, however, until now there has not been a single Cepheid with a
purely dynamical measurement of its mass.  

Polaris ($\alpha$ UMi) is the nearest and, at 2nd magnitude, the brightest
classical Cepheid, albeit one with a small light amplitude in its 3.97-day
pulsation period (Turner et al.\ 2005 and references therein). The amplitude,
which had been decreasing for several decades, now appears to have stabilized
and may be increasing again (Bruntt et al.\ 2008 and references therein). The
{\it Hipparcos\/} parallax of Polaris indicates a luminosity consistent with
pulsation in the first overtone (Feast \& Catchpole 1997).

Polaris is the brightest member of a triple system (see Kamper 1996 and
references therein). The well-known visual companion, Polaris~B, is an
8th-magnitude F3~V star at a separation of $18''$.  The Cepheid itself is a
member of a single-lined spectroscopic binary with a period of 30~yr. In this
paper we report the first direct detection of the close companion, from which we
derive the first entirely dynamical mass measurement for a Cepheid.

Cepheid masses are a key parameter for testing stellar evolutionary
calculations.  Beginning in the 1960's, discrepancies were found in the sense
that Cepheid masses derived from pulsation modeling were lower than those
derived from evolutionary tracks.  A revision in envelope opacities brought
pulsation and evolutionary masses closer together, partially alleviating this
``Cepheid mass problem.''  However, recent evolutionary and pulsation
constraints for Galactic (Bono et al.\ 2001b; Caputo et al.\ 2005; Keller 2008;
Natale, Marconi, \& Bono 2008) and Magellanic (Bono, Castellani, \& Marconi
2002; Keller \& Wood 2006) Cepheids still imply a discrepancy in masses at the
$\sim$15-20\% level.  The luminosities of Cepheids depend on the mass of the
helium-burning core; physical mechanisms affecting the helium core mass include
mixing due to convective core overshoot during the main-sequence phase, mass
loss, stellar rotation, and radiative opacity. A directly measured mass for
Polaris would thus provide an important constraint on this theoretical
framework. 

\section{Observations and Data Reduction}

With the intention of a direct detection of the close companion, we imaged
Polaris with the {\it Hubble Space Telescope\/} ({\it HST\/}) and the High
Resolution Channel (HRC; plate scale $0\farcs026 \,\rm pixel^{-1}$) of the
Advanced Camera for Surveys (ACS)\null. We chose the ultraviolet F220W filter
(effective wavelength $\sim$2255~\AA) in order to minimize the contrast between
Polaris and the close companion, which we anticipated to be a main-sequence star
slightly hotter than the Cepheid, and also to minimize the size of the
point-spread function (PSF)\null. 

Observations were obtained on 2005 August 2-3, and again on 2006 August 13. At
the first epoch we obtained a series of 0.1 to 0.3~s exposures dithered across
200 pixels on the chip over the course of one {\it HST\/} orbit, with several
exposures taken at each dither position.  At the second epoch we used the same
dither pattern, but divided the spacecraft orbit between a series of 0.3~s
exposures on Polaris~A and 20~s exposures on Polaris~B\null.  For the longer
exposures, we placed Polaris~B at the same chip location as Polaris~A in the
short exposures, so as to provide an accurate PSF for a single star at the same
place in the field of view.

Figure 1 shows the co-added images of Polaris~A from 2005 (left panel) and 2006
(middle panel).  The close companion (which we designate Polaris~Ab) is 
detected at the lower left of the primary (at about a ``7 o'clock'' position).
Because of the asymmetric PSF shape, we performed several checks to confirm that
the apparent companion is not an artifact. The right-hand frame in Figure~1
shows Polaris~B in the 2006 image, with the star shifted to the same field
position as Polaris~A, and with its image scaled to the same flux as
Polaris~A\null. This PSF shows no artifact at the location of the companion seen
in the images of Polaris~A. 

Figure 2 shows contour maps of the same three images. Again the faint companion
is seen in the 2005 and 2006 images of Polaris~A, and there is no PSF artifact
at this location in the image of Polaris~B\null. We also retrieved images from
the {\it HST\/} archive of several standard stars observed with ACS/HRC in
the F220W filter over an interval of four years.  Examples of these observations
are shown in Figure~3.  Although the PSF structure does vary somewhat with
time, due to small changes in telescope focus and other instrumental phenomena,
none of these images show any artifact at the location of the Polaris Ab
companion seen in Figures 1 and 2.

To measure the separation and position angle between Polaris Aa and the newly
revealed close companion, we used the calibrated flat-fielded exposures provided
by the {\it HST\/} reduction pipeline. At each dither location, we
median-filtered the repeated observations to remove cosmic rays.  We then
extracted subarrays from the images centered on Polaris Aa with a size of
$0\farcs85\times0\farcs85$.  We used the observations of Polaris~B from 2006 as
a reference PSF to construct models of the close pair (Aa,Ab) by searching
through a grid of separations and flux ratios. The IDL {\it interpolate\/}
procedure was used to shift the PSF by sub-pixel intervals, using cubic
convolution interpolation. The background was least-squares fitted with a tilted
plane.

The separations, position angles (PAs), and magnitude differences, determined
through $\chi^2$ minimization between the models and the observations, are given
in Table~1.  The uncertainties were determined by analyzing multiple images
individually and computing the standard deviation. We applied the
filter-dependent geometric distortion correction of Anderson \& King (2004) to
convert the pixel values to a separation in arcseconds. To define the
orientation of the detector $y$-axis on the sky, and thus determine the PA of
the binary relative to the pole for the equinox J2000.0, we used the {\it HST\/}
image-header keyword PA\_APER\null.  

We also measured the separation and PA of the wide companion, Polaris~B,
relative to Aa, and the results are presented in Table~2.  The good agreement
with the historical measurements of the PA of Polaris B relative to A (see \S 4)
indicates that we are properly defining the direction of north in spite of the
extreme northerly declination. (Since the historical double-star convention is
to give the PA for the equinox of the date of observation, we computed the
precession corrections and give the adjusted PAs in the footnotes to Tables~1
and~2, for the convenience of archivists.)

\section{Orbital Solutions}                              
   
We stress that the orbital analyses discussed below are only preliminary fits to
data with a very limited sample (only two points) of separations and PAs. We
followed three different approaches to determining the orbital parameters, in
order to illustrate the scope of the available data.  

Kamper (1996) rederived the single-lined spectroscopic orbit of Polaris~Aa using
improved radial-velocity data, and a careful removal of the velocity signal due
to the Cepheid pulsation. His solution provides the period, time of periastron
passage, eccentricity, angle between the node and periastron, and the
radial-velocity semi-amplitude of the primary (denoted $P, T, e, \omega,$ and
$K_{\rm Aa}$, respectively). By comparing the {\it Hipparcos\/} proper motion of
Polaris~Aa (which, over the duration of the {\it Hipparcos\/} mission, is nearly
instantaneous in the context of the 30-year orbit) with the ground-based
long-term average proper motion from the FK5 (which is essentially the motion of
the center of mass), Wielen et al.\ (2000) determined the inclination and the PA
of the line of nodes ($i$ and $\Omega$). Their analysis, however, allows for
retrograde and prograde orbital solutions (the two orbits being tangential at
the {\it Hipparcos\/} epoch), with two different values of $i$ and $\Omega$. The
orbital parameters based on the Kamper (1996) and the two Wielen et al.\ (2000)
solutions are presented in Table~3. 

The {\it HST\/} detection of the close companion Ab and its orbital motion at
two epochs establishes a retrograde sense for the orbit (thus confirming the
strong preference stated by Wielen et al.\ for their retrograde solution).
Additionally, it provides constraints on the remaining unknown parameter of the
orbit, the semi-major axis $a$.  A combination of the spectroscopic mass
function,
\begin{displaymath} 
f(M) = (M_{\rm Ab}\sin{i})^3/(M_{\rm Aa} + M_{\rm Ab})^2 = 
 3.784\times10^{-5} K_{\rm Aa}^3 P (1 - e^2)^{3/2} \, ,
\end{displaymath} 
with the total mass from Kepler's Third Law,
\begin{equation}
M_{\rm tot} = M_{\rm Aa} + M_{\rm Ab} = \frac{a^3}{\pi^3P^2} \, ,    
\end{equation}  
then yields the masses of the binary components,
\begin{displaymath} 
M_{\rm Ab} = 0.03357 \frac{K_{\rm Aa} a^2 \sqrt{1 - e^2} }{\pi^2 P \sin{i} } 
  \, , 
\end{displaymath} 
\begin{displaymath} 
M_{\rm Aa} = M_{\rm tot} - M_{\rm Ab} \, ,
\end{displaymath} 
where $a$ and the parallax $\pi$ are in arcseconds, $P$ is in years,
$K_{\rm Aa}$ is in km~s$^{-1}$, and the masses are in $M_\odot$. 

In \S3.1-3.3 we describe the orbital fits that we computed based on a synthesis
of the spectroscopic, astrometric, and {\it HST\/} data.  In these sections, we 
present three successively more comprehensive orbital solutions. Table~4
summarizes the orbital parameters determined from each of these fits; the
individual columns are described in more detail in \S3.1-3.3. 

\subsection{Semi-major Axis} 

As a first approximation to an orbital solution, we fixed the spectroscopic and
astrometric parameters ($P, T, e, \omega, i, \Omega$) to be those determined by
Kamper (1996) and Wielen et al.\ (2000), and solved only for the semi-major axis
$a$ based on the two {\it HST\/} separation measurements.  The orbital
parameters for this solution are listed in the second column of Table~4. 
Figure~4 compares this retrograde orbit fit with the {\it HST\/} measurements,
and shows extremely poor agreement for the PAs. However, the relatively large
uncertainties in $i$ and $\Omega$ determined by Wielen et al.\ (2000) provide
considerable flexibility for adjusting the orbital parameters in order to
improve the fit quality.

\subsection{Best Fit to the {\it HST\/} Measurements}

To get a better fit to the {\it HST\/} data, we solved for $i$, $\Omega$, and
$a$ based on the two separation and PA measurements from the {\it HST\/}
observations, while holding the relatively well-determined spectroscopic
parameters ($P, T, e,$ and $\omega$) fixed. We computed the orbit fit through a
standard Newton-Raphson method in $\chi^2$ space, and present the results in the
third column of Table~4. 

To explore the range of orbital parameters that fit the {\it HST\/} data, we
performed a Monte Carlo search by selecting values of $i$, $\Omega$, and $a$ at
random. We searched for 10,000 solutions within the 3$\sigma$ confidence
interval, corresponding to a difference of $\Delta\chi^2 = 9$ from the minimum
$\chi^2$ value. Figure 5 shows cross-cuts through the $\chi^2$ surfaces for the
three derived parameters. Using the recently revised {\it Hipparcos\/} parallax
of $7.72 \pm 0.12$~mas (van~Leeuwen et al.\ 2007), we computed the total mass of
the binary through Kepler's Third Law for all of the solutions found in the
Monte Carlo search.  In the last panel of Figure 5 we show a plot of the total
mass versus inclination. Because a visual orbit is insensitive to the individual
masses of the components, when combining the total mass with the spectroscopic
mass function, there exist values of the inclination that produce negative
masses for $M_{\rm Aa}$ or $M_{\rm Ab}$.  In the remaining analysis we removed
these negative-mass solutions from our sample of possible orbits. Essentially,
this rejects all orbital solutions with $i > 168^\circ$. The 1$\sigma$
uncertainties listed in the third column of Table~4 are determined from the
$\Delta\chi^2 = 1$ confidence interval of the modified distribution. The values
of $i$ and $\Omega$ agree with the retrograde parameters computed by Wielen et
al.\ (2000) at the 1.5$\sigma$ and 2.2$\sigma$ levels respectively. Figure 6
shows three examples illustrating how the orbit fit varies within a 1$\sigma$
confidence interval.  

We note that the {\it HST}-only solution yields a secondary mass of $M_{\rm Ab}
= 1.8 \, M_\odot$ (Table~4, col.~3), corresponding approximately to an A5 V
star.  The discussion below in \S5.1, as well as the lack of a detection of the
companion in UV spectra obtained by Evans (1988) with the {\it International
Ultraviolet Explorer\/} ({\it IUE\/)}, makes it highly improbable that Ab could
be this hot. With only two {\it HST\/} measurements sampling the orbit thus far,
we do not yet have a good constraint on the curvature, and hence acceleration,
of the orbit. In turn, this limits how well we can determine the total system
mass, and it contributes to the large errors quoted in Table~4.  

\subsection{Joint Fit to {\it HST\/} and Proper-Motion Measurements}

Incorporation of the {\it Hipparcos\/} proper-motion measurements into the
orbital fit extends the time coverage of the measurements to $\sim$15 years. 
This represents a significant fraction of the orbital period and is therefore
likely to improve the reliability of the results.  Following the technique
described in Wielen et al.\ (2000), we performed a simultaneous fit to the
proper-motion data and the {\it HST\/} measurements.  

As Wielen et al.\ point out, the FK5 proper motion is averaged over several
cycles of the orbital period and therefore reflects the center-of-mass motion of
Polaris~Aa,Ab. Because of the shorter time span of the {\it Hipparcos\/}
mission, the {\it Hipparcos\/} proper motion more nearly represents an
instantaneous measurement of the combined proper motion of the center of mass of
the Aa,Ab pair and the orbital motion of the photocenter about the center of
mass at the epoch of the observations ($\sim$1991.3).  The difference between
the FK5 and {\it Hipparcos\/} proper motions thus gives the offset caused by the
orbital motion.  

In computing the joint fit, we held the spectroscopic parameters ($P, T, e,
\omega$, and $K_{\rm Aa}$) fixed at the Kamper (1996) values and solved for 
$i$, $\Omega$, and $a$, again using a Newton-Raphson technique in $\chi^2$
space. The input data were the relative positions of Aa and Ab at the two {\it
HST\/} epochs, and the difference between the {\it Hipparcos\/} and FK5 proper
motions. To incorporate the proper-motion data into the orbit fit, we had to
compute the time-dependent offset of the photocenter relative to the center of
mass predicted by the orbital parameters during the time of the {\it
Hipparcos\/} observations. To compute these offsets, we converted the semi-major
axis of Polaris~Aa determined from the single-lined spectroscopic orbit to the 
semi-major axis of the photocenter by using the mass ratio computed from the
full set of orbital parameters and a magnitude difference between Aa and Ab of
$\Delta V = 7.2$ (see \S5.1).  This conversion is specified by eqns.~(8) and
(9) in Wielen et al.\ (2000).  The expected difference between the
instantaneous and the mean proper motion ($\Delta \mu_{\rm fit}$) at the
central epoch of the {\it Hipparcos\/} observations is then computed from
eqn.~(18) of Wielen et al.

Table 5 shows the values of the proper motions used by Wielen et al.\ (2000). 
The first line shows the proper motion of Polaris given in the FK5 catalog
($\mu_{\rm FK5}$).  In the second line, a systematic correction is applied to
convert the proper motion from the FK5 reference system to the {\it
Hipparcos\/}/ICRS system (Wielen et al.\ 2000).  The proper motion measured by
{\it Hipparcos\/} ($\mu_{HIP}$) is given in the third line.  The difference in
the measured proper motions $\Delta\mu = \mu_{HIP} - \mu_{\rm FK5}$ is given in
the fourth line.  

The last line in Table~5 shows the best-fit difference between the 
instantaneous and mean proper motion ($\Delta\mu_{\rm fit}$) calculated from our
simultaneous fit to the relative separation and PA measurements of Polaris Aa,Ab
and the proper-motion data. Figure~7 shows a graphical representation of the
{\it Hipparcos\/} and FK5 proper motions, and compares our best-fit value with
the measured value for the difference between the instantaneous and mean proper
motions.

As Figure~7 illustrates, the best-fit value of $\Delta\mu$ in right ascension
agrees well ($0.7\sigma$) with the measured value, but the agreement is poorer
($3.2\sigma$) in declination. This discrepancy probably arises from our
constraining the spectroscopic parameters to be exactly those derived by Kamper
(1996), thus forcing the proper motions and {\it HST\/} measurements to absorb
the errors. We found that by allowing some variation in the spectroscopic
parameters, we could substantially improve the fit to the proper-motion
measurements. Once we have sampled enough of the visual orbit to better
constrain $i$, $\Omega$, and $a$, the optimal orbital solution should be found
by doing a simultaneous fit to the radial-velocity, {\it HST\/}, and
proper-motion data. Unfortunately, however, Kamper only tabulated the radial
velocities {\it before\/} removal of the pulsational variation, so a
re-computation of the pulsation corrections would have to be carried out---a
task well beyond the scope of this paper, and one that should await the
availability of more {\it HST\/} observations.

The orbital parameters and derived masses from our final combined fit are given
in the last column of Table~4, which also contains our final best estimates for
the dynamical masses of both stars. The best-fit orbit of Polaris Ab relative to
Aa is plotted in Figure~8. The shaded-gray portion of the orbit marks the
location of the companion during the interval of the {\it Hipparcos\/}
observations in the early 1990's, and it should be noted that the direction of
motion at that time was, of course, $180^\circ$ different from the direction of
the differential motion of Polaris~Aa ($\Delta\mu_{\rm fit}$) shown as a green
arrow in Figure~7.

\section{Astrometry of Polaris B}

Visual measurements of the PA and separation of the wide companion Polaris~B
relative to Polaris~A extend back to the early 19th Century, with a few
photographic observations being available from the 20th Century. We have
compiled these measurements from Kamper (1996) and the Washington Double Star
Catalog (Mason et al.\ 2001)\footnote{See http://ad.usno.navy.mil/wds}. 

In Figure~9 we plot the PA measurements of Polaris~B, which have been precessed
to the J2000.0 equinox. Due to its slow relative motion and large magnitude
difference, Polaris A,B was generally not included in the major double-star
observing programs, especially in the latter half of the 20th Century. This had
the consequence that most of the measures that were obtained tended to be made
by less-experienced observers, often using older techniques and/or smaller
telescopes; this may explain the surprisingly large scatter in the late 20th
Century. A linear least-squares fit to the data yields a rate of change in the
PA of  $-0\fdg00035 \pm 0\fdg00094 {\rm ~yr^{-1}}$, consistent with no
detectable change in PA for the past two centuries. (The earlier work of Kamper
had given a marginal detection of $+0\fdg0086 \pm 0\fdg0076 {\rm ~yr^{-1}}$.)  

The separation measurements for Polaris~B are plotted in Figure~10.  There has
been a slow downward trend in the separation, with a least-squares fit giving a
rate of $-1.67\pm0.19$~mas~yr$^{-1}$. Kamper (1996), from the ground-based
measures only, had found $-1.7\pm0.6$~mas~yr$^{-1}$.  Since the absolute proper
motion of Polaris~A is $\sim$46~mas~yr$^{-1}$ (see Table~5), the absolute
motions of A and B agree to within about 4\%.  At the distance of Polaris given
by the {\it Hipparcos\/} parallax, the difference in tangential velocities
between A and~B is $1.03\pm0.12 \,\rm km \, s^{-1}$. 

In computing these least-squares fits, we weighted the observations by estimates
of their measurement errors.  For the {\it HST\/} observations, we applied the 
uncertainties quoted in Table~2.  For the historical measurements, we divided 
the data into four groups spanning approximately 50 years each. We assumed
measurement uncertainties equal to the standard deviation of the values
measured in each of these four groups.

The slowly diminishing separation of A and B (at constant PA) is not
inconsistent with orbital motion in a physically bound pair---which is also
supported by the close similarities of the radial velocities of A and B (Kamper
1996; Usenko \& Klochkova 2007).  To predict an order-of-magnitude rate of
change in the separation, we assumed a circular orbit with a total system mass
of $M_{\rm Aa} + M_{\rm Ab} + M_{\rm B} = 7.15 \, M_\odot$ (based on the Aa+Ab
mass of $5.8\,M_\odot$ in the last column of Table~4 and a mass for Polaris~B of
$1.35\,M_\odot$---see below). Adopting the revised {\it Hipparcos\/} parallax of
7.72~mas, and assuming an edge-on orbit with a period of $\sim$100,000 yr, we
find a semimajor axis of $a\simeq32''$ (or 0.02~pc). At the orbital phase
implied by the observed separation of $18\farcs2$, the relative motion would
then be $-1.65$~mas~yr$^{-1}$, close to the observed value.

\section{Astrophysical Properties of the Companions of Polaris}

\subsection{Polaris Ab}

Our observed ultraviolet magnitude difference between Polaris~Aa and Ab may be
used to infer the spectral type, and hence the mass, of the newly resolved close
companion.  

We downloaded UV spectra from the {\it IUE\/} data archive\footnote{The {\it
IUE\/} data were obtained from the Multimission Archive at the Space Telescope
Science Institute (MAST)\null.  Support for MAST for non-{\it HST\/} data is
provided by the NASA Office of Space Science via grant NAG5-7584 and by other
grants and contracts.} for three F-type dwarfs having accurate spectral types
and parallaxes, as well as for Polaris itself. The F~stars selected were 78~UMa
(HR~4931, HD~113139; F2~V), HD~27524 (F5~V), and HD~27808 (F8~V). The latter two
stars (Hyades members), as well as Polaris itself, were taken to be unreddened,
while the spectrum of 78~UMa (a member of the Ursa Major group) was dereddened
by $E(B-V) = 0.01$~mag.  We then scaled the flux distributions for the three
stars to the distance of Polaris, using the respective {\it Hipparcos\/}
parallaxes.

In Table~6, we list these comparison stars, their spectral types, masses implied
by the spectral types, their {\it Hipparcos\/} parallaxes, absolute magnitudes
based on the parallaxes, and finally the predicted flux ratios relative to
Polaris in the ACS/HRC F220W band. The adopted relationship between spectral
types and masses is that of Harmanec (1988).  The flux ratios were calculated by
convolving the F220W system-throughput function (see Chiaberge \& Sirianni 2007)
with the scaled {\it IUE\/} spectra, and then ratioing with respect to Polaris.
In Figure~11 we show the {\it IUE\/} spectrum of Polaris and the scaled spectra
of the three F dwarfs.
    
As listed in Table~1, the observed magnitude difference in the ACS/HRC F220W
filter between Polaris~Aa and Ab is $5.39\pm0.08$~mag, or a flux ratio of
$0.0070 \pm 0.0006$. A small correction to this ratio is needed because of the
small ($\sim$10-15\%) contribution to the signal from the red leak in the F220W
filter, Polaris~Aa being slightly redder than the companion. The red-leak
contributions have been tabulated as a function of spectral type by Chiaberge \&
Sirianni (2007), leading to a corrected in-band flux ratio of $0.0074 \pm
0.0006$. This value is entered in the fourth row of Table~6, and is marked with
a horizontal line in Figure~11. Interpolation in the last column of Table~6 then
leads to an inferred spectral type of about F6~V, an absolute visual magnitude
of $M_V\simeq+3.6$, and an expected mass of $1.3\,M_\odot$\null.

The apparent $V$-band magnitude of Polaris~Ab, for a distance modulus
$(m-M)_0=5.56$, is inferred to be about 9.2, or some 7.2~mag fainter at $V$ than
Polaris~Aa.  This illustrates the advantage of observing the Polaris system in
the ultraviolet, which lessens the contrast by nearly 2~mag. 

\subsection{Polaris B}

As listed in Table~2, we also measured the F220W flux difference between
Polaris~Aa and~B as $4.49\pm0.04$~mag, or a flux ratio of $0.0160 \pm 0.0006$. 
Correction for red leak, as described above, changes the flux ratio to $0.0178
\pm 0.0006$, entered in the second row of Table~6, and also marked with a
horizontal line in Figure~11. Interpolating again in the table, we see that this
ratio corresponds to a star intermediate between types F3~V and F4~V\null. In
the case of the well-resolved Polaris~B the optical spectral type has been
determined from the ground. Our result is in gratifying agreement with the
spectral type of F3~V found by Turner (1977) and Usenko \& Klochkova (2007), who
also cite earlier spectral classifications of Polaris~B by experts such as
Bidelman. This finding not only validates our photometric analysis of Polaris~Ab
above, but again supports the physical association of Polaris~A and~B\null. 
Based on the relationship between spectral type and mass in Table~6, we infer
the mass of Polaris~B to be near $1.35\,M_\odot$.

Using the same method as for Polaris~Ab, we can use the UV flux ratio to infer
the $V$ magnitude of Polaris~B to be 8.7. The visual magnitude of Polaris~B can
be measured from the ground, but is made difficult by scattered light from
Polaris~A\null.  Kamper (1996) used CCD imaging to determine a magnitude
difference with respect to A of $\Delta V = 6.61$, implying $V=8.59$ (in good
agreement with earlier photoelectric measurements of $V=8.5$ and 8.60 by Fernie
1966 and McNamara 1969, respectively; Fernie included an approximate correction
for scattered light, and McNamara states that he observed only on excellent
nights). In more recent work, to be reported separately, we have been carrying
out astrometry of Polaris~B with the Fine Guidance Sensors (FGS) onboard {\it
HST\/}\null. As a byproduct, these observations yield an accurate $V$ magnitude
of $8.65\pm0.02$.  Thus our indirectly inferred $V$ magnitude for Polaris~B of
8.7 agrees very well with the ground- and {\it HST-}based observations. 

\section{Dynamical Masses}

\subsection{Polaris Ab}

The final column in Table~4 lists the dynamical masses of both components of the
close pair Aa,Ab obtained from our final orbital solution, as described in
\S3.3. For Ab, the dynamical mass is $1.26^{+0.14}_{-0.07}\,M_\odot$\null. This
is in remarkably good agreement with the $1.3\,M_\odot$ inferred indirectly from
the UV flux difference (\S5.1), and is an indicator of the validity of our
orbital solution.

\subsection{Theoretical Implications of the Cepheid's Dynamical Mass}

The dynamical mass of the Cepheid Polaris~Aa from our final orbital solution, as
listed in the last column in Table~4, is $4.5^{+2.2}_{-1.4} \, M_\odot$\null. 

We compare this result first with theoretical ``evolutionary'' masses,
$M_e$\null.  The input data are the intensity-averaged mean apparent magnitudes
($m_V=1.98$, $m_B=2.58$, from Fernie et al.\ 1995 and assumed to be unreddened),
the revised {\it Hipparcos\/} distance of $129.5\pm2.0$~pc (van~Leeuwen et al.\
2007), and a solar metal abundance (Luck \& Bond 1986; Usenko et al.\ 2005). We
adopt the mass-period-luminosity (MPL) relation for He-burning fundamental
pulsators provided by Caputo et al.\ (2005, their Table~4). Before using this
relation, we fundamentalized the first-overtone (FO) pulsation period of Polaris
with the relationship $\log P_{\rm F} = \log P_{\rm FO} + 0.13$. By assuming
Cepheid luminosities predicted by ``canonical'' evolutionary models that neglect
convective-core overshooting, we find $M_e=6.1\pm 0.4\, M_\odot$\null.  On the
other hand, if we assume luminosities predicted by ``noncanonical'' evolutionary
models that account for mild convective-core overshooting, given by $L/L_{\rm
can}\simeq 1.3$, we find $M_e=5.6\pm 0.4\, M_\odot$\null.  Use of the
mass-color-luminosity (MCL) relation (Caputo et al.\ 2005, Table~5) yields very
similar evolutionary masses. 

The HR diagrams in Figure~12 show a direct comparison between theoretical
evolutionary tracks and observations in the $M_V, (B-V)_0$ plane. In both panels
of Figure~12 we plot the location of Polaris with an open triangle enclosing a
small error bar. The top panel shows canonical evolutionary tracks at solar
chemical composition; $M_e \simeq 6\, M_\odot$ provides a good fit to the
position of Polaris. The bottom panel shows noncanonical tracks, suggesting $M_e
\simeq 5.5\, M_\odot$, except that the tip of the blue loop is not quite as hot
as Polaris. However, the blueward extension of the loops is affected by chemical
composition and by physical and numerical assumptions (Stothers \& Chin 1991;
Chiosi, Bertelli, \& Bressan 1992; Bono et al.\ 2000; Meynet \& Maeder 2000; Xu
\& Li 2004).  

To compare our result with ``pulsation'' masses, $M_p$, we used the
mass-dependent period-luminosity-color (PLC) relation of Caputo et al.\ (2005,
their Table~2). We again fundamentalized the pulsation period of Polaris, and
using its intensity-averaged value of $M_V$ find $M_p{\rm(PLC)}=5.1\pm 0.4 \,
M_\odot$\null.   The pulsation mass of Polaris can also be estimated using the
predicted period-mass-radius (PMR) relation for first-overtone Cepheids of Bono 
et al.\ (2001a), along with the radius of Polaris, $R=46\pm3\, R_\odot$, 
measured interferometrically by Nordgren et al.\ (2000). This gives 
$M_p{\rm(PMR)}=4.9\pm 0.4 \, M_\odot$. 

The lower pulsation masses, taken at face value, are thus in better agreement
with the nominal dynamical mass than are the higher evolutionary masses.
However, the current $1\sigma$ range of the measured dynamical mass,
3.1--$6.7\,M_\odot$, encompasses the entire range of theoretical masses. Thus,
our discussion serves mainly to emphasize the crucial importance of reducing the
error bars through continued {\it HST\/} high-resolution imaging of the Polaris
Aa,Ab system. In addition, our companion {\it HST\/} FGS astrometric program
will provide an improved trigonometric parallax. Doubtless there will also be
future improvements in the spectroscopic orbit (e.g., Turner et al.\ 2006;
Bruntt et al.\ 2008).  Simulations suggest that we can reduce the uncertainty on
the Cepheid's dynamical mass to below $\pm$$0.6\,M_\odot$\null. This would
provide a major constraint on the evolution of intermediate-mass stars and the
physics of Cepheid pulsation.

\subsection{Issues in the Evolution of Intermediate-Mass Stars}

As a further elaboration of the importance of accurate dynamical masses for
Cepheid variables, we summarize the major open questions in the calculation of
evolutionary tracks of intermediate-mass stars. A more complete discussion is
provided by Bono, Caputo, \& Castellani (2006).

The luminosity of an evolved intermediate-mass star is related to the mass of
the He-burning core. The physical mechanisms affecting the core mass include:

\begin{enumerate}

\item ``Extra-mixing'' of hydrogen into the core through convective core
overshooting during the central hydrogen-burning phases; 

\item Mass loss, leading to a lower total stellar mass at the same luminosity;

\item Rotation: the shear layer located at the interface between the convective
and radiative regions enhances internal mixing, producing a larger He core mass;

\item Radiative opacity: an increase in stellar opacity causes an increase in
the central temperature, enhanced efficiency of central H-burning, and a higher
core mass.

\end{enumerate}

Here we briefly discuss a few recent results that bear on convective overshoot
and mass loss.

The discussion of the Polaris mass in the previous subsection showed that the
inclusion of noncanonical overshoot gave a better agreement with our preliminary
dynamical mass. This is borne out by a mass measurement for the longer-period
Cepheid S~Muscae, based on {\it HST\/} Goddard High Resolution Spectrograph
radial velocities of its hot companion, and an assumed companion mass
based on its {\it FUSE\/} spectrum (Evans et al.\ 2006). The implied mass of
S~Mus clearly favors mild convective overshoot.  

For mass loss, we note that the evolutionary calculations discussed in the
previous subsection included semi-empirical mass-loss rates (Reimers 1975; 
Nieuwenhuijzen \& de Jager 1990). These rates are insufficient to resolve the
discrepancy between evolutionary and pulsational masses. However, a variety of
mostly recent observational information suggests that mass loss from Cepheids
may be significant. At least two Cepheids, SU~Cas and RS~Pup, are associated
with optical reflection nebulae (see Kervella et al.\ 2008 and references
therein) that may represent mass ejection from the Cepheids. Moreover, a large
circumstellar envelope around the Cepheid $\ell$~Car has been detected recently
by Kervella et al.\ (2006), using mid-infrared data collected with the MIDI
instrument on the VLTI\null.  Circumstellar material has also been detected
around $\delta$~Cephei and Polaris itself by M\'erand et al.\ (2006).  Recent
{\it Spitzer\/} observations of a sample of Cepheids (Evans et al.\ 2007) have
likewise revealed an infrared excess in the direction of $\delta$~Cephei. 

On the theoretical side, a recent investigation (Neilson \& Lester 2008)
indicated that the coupling between radiative line driving (Castor et al.\ 1975)
and the momentum input of both radial pulsation and shocks can provide mass-loss
rates for Galactic Cepheids ranging from $10^{-10}$ to $10^{-7} \, M_\odot\,\rm
yr^{-1}$. This finding, together with typical evolutionary lifetimes (e.g., Bono
et al.\ 2000, Table~7), indicates that classical Cepheids may in fact be capable
of losing the 10-20\% of their mass that would be needed to resolve the
discrepancy.

\section{Summary}

The results of this study are as follows:

\begin{enumerate}

\item We have used UV imaging with the ACS/HRC onboard the {\it HST\/} to make
the first direct detection of the close companion of the classical Cepheid
Polaris.

\item We confirm orbital motion in a retrograde sense, based on two observations
a year apart.

\item By combining our {\it HST\/} measurements with the single-lined
spectroscopic orbit (Kamper 1996) and the FK5 and {\it Hipparcos\/} proper
motions (Wielen et al.\ 2000), we derive a dynamical mass for the Cepheid
Polaris~Aa of $4.5^{+2.2}_{-1.4}\,M_\odot$---the first purely dynamical mass for
any Cepheid.

\item The dynamical mass is smaller than values estimated either from
pulsational properties or evolutionary tracks, but the error bars are still
large enough that the discrepancies have not achieved statistical significance.

\item The close companion Polaris Ab has a dynamical mass of
$1.26^{+0.14}_{-0.07}\,M_\odot$. This is consistent with a spectral type of
about F6~V, inferred from the UV brightness of Ab.

\item The more distant and well-known companion Polaris~B has a UV flux
consistent with its known spectral type of F3~V, lying at the same distance as
the Cepheid. The proper motion of Polaris~B is shown to be very similar to that
of Aa,Ab, consistent with motion in a wide but bound orbit around the close
pair.

\item Continued {\it HST\/} imaging, including two more observations that have
been approved for our own program, will decrease the errors on the dynamical
mass of Polaris, allowing a critical test of stellar-evolution theory and the
influence of such effects as convective overshoot, mass loss, rotation, and
opacities.

\end{enumerate}

\acknowledgments

We are happy to acknowledge financial support from STScI grants GO-10593,
GO-10891, and GO-11293 (NRE and HEB), and Chandra X-ray Center NASA Contract
NAS8-03060 (NRE and MK)\null. This research has made use of the Washington
Double Star Catalog maintained at the U.S. Naval Observatory. The contributions
of the late Karl Kamper to the study of Polaris were crucial to this work.

\clearpage       
                 
\begin{figure}
\plotone{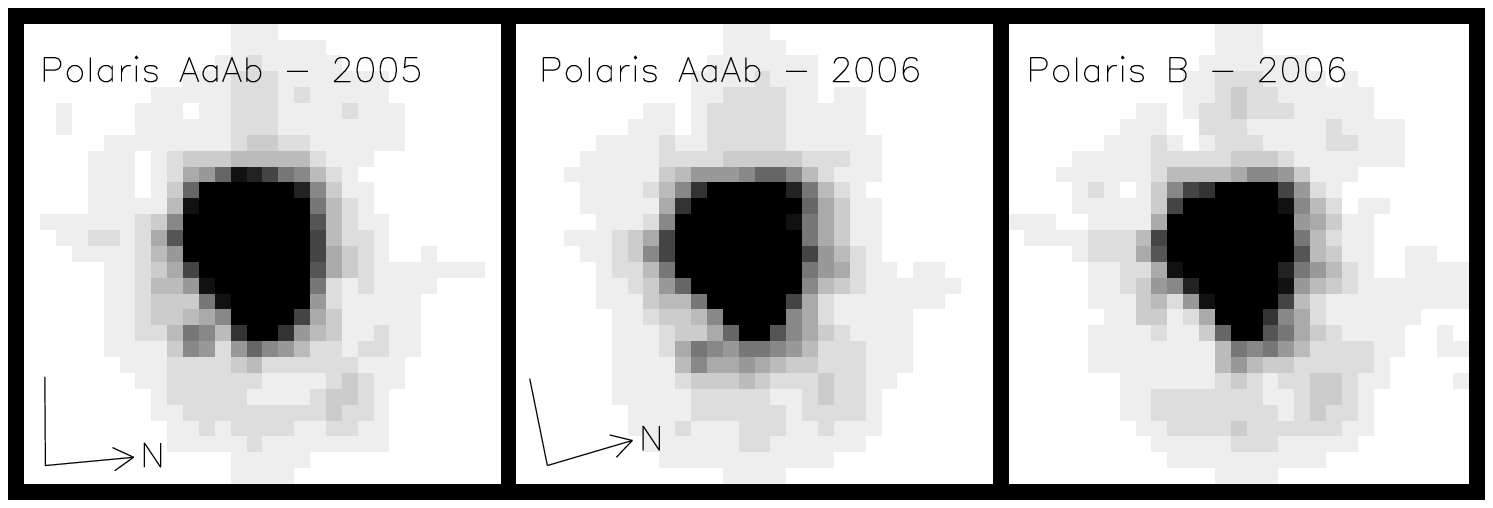}  

\caption{Coadded ACS HRC images of Polaris Aa,Ab taken with the F220W filter on
2005 Aug 2 (left) and 2006 Aug 13 (middle). The close companion Ab is detected
at the lower left of the primary (at about a ``7 o'clock'' position). The images
are $0\farcs85\times0\farcs85$ and the directions of N and E are indicated. The 
right-hand panel shows a coadded image of Polaris~B from longer exposures taken
during the 2006 observations, and scaled to the flux level of the Polaris Aa,Ab
images. There is no artifact in the Polaris~B PSF at the location of Ab.}

\end{figure}

\begin{figure}
\plotone{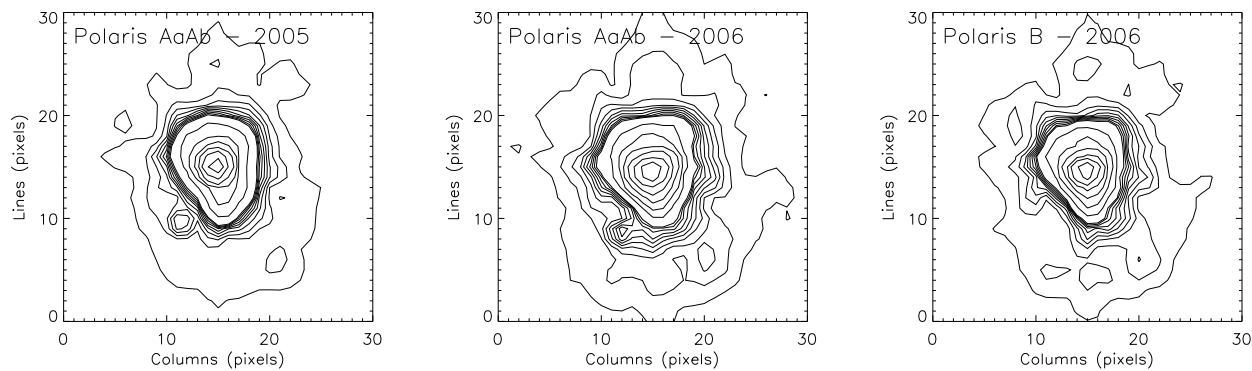}  
\caption{Contour maps of the coadded images shown in Figure~1. The outermost
contour intervals in each panel range from 0.001 to 0.010 of the peak flux in
steps of 0.001, and thereafter are at levels of 0.02, 0.04, 0.08, 0.16, 0.32,
and 0.64 of the peak flux. The contours again demonstrate the absence of any
artifact at the location of the Ab companion.}
\end{figure}

\begin{figure}
\plotone{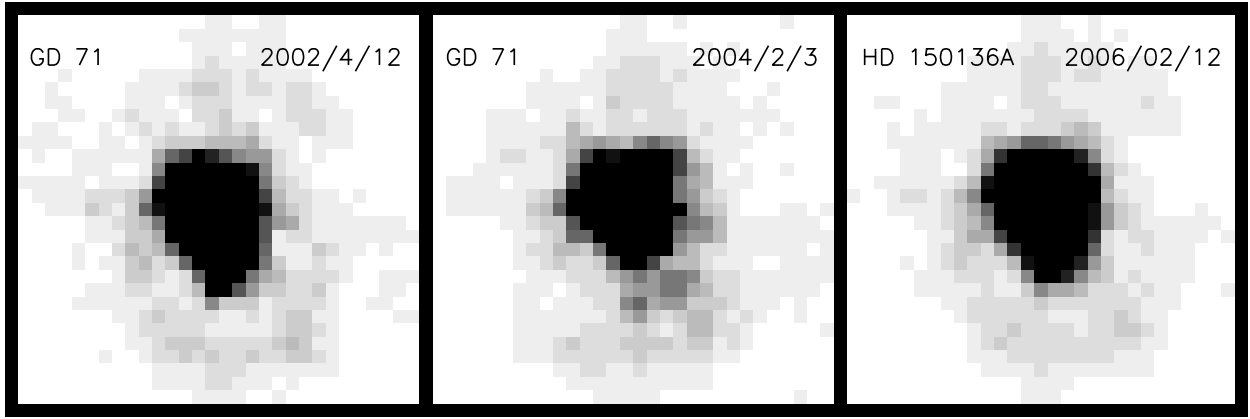}  
\caption{Archival images of flux-calibration standard stars observed with the
ACS/HRC in the F220W filter. The star names and dates of observation are
listed in each panel. There is no artifact at the location of the Polaris Ab
companion.}
\end{figure}

\begin{figure}
\plotone{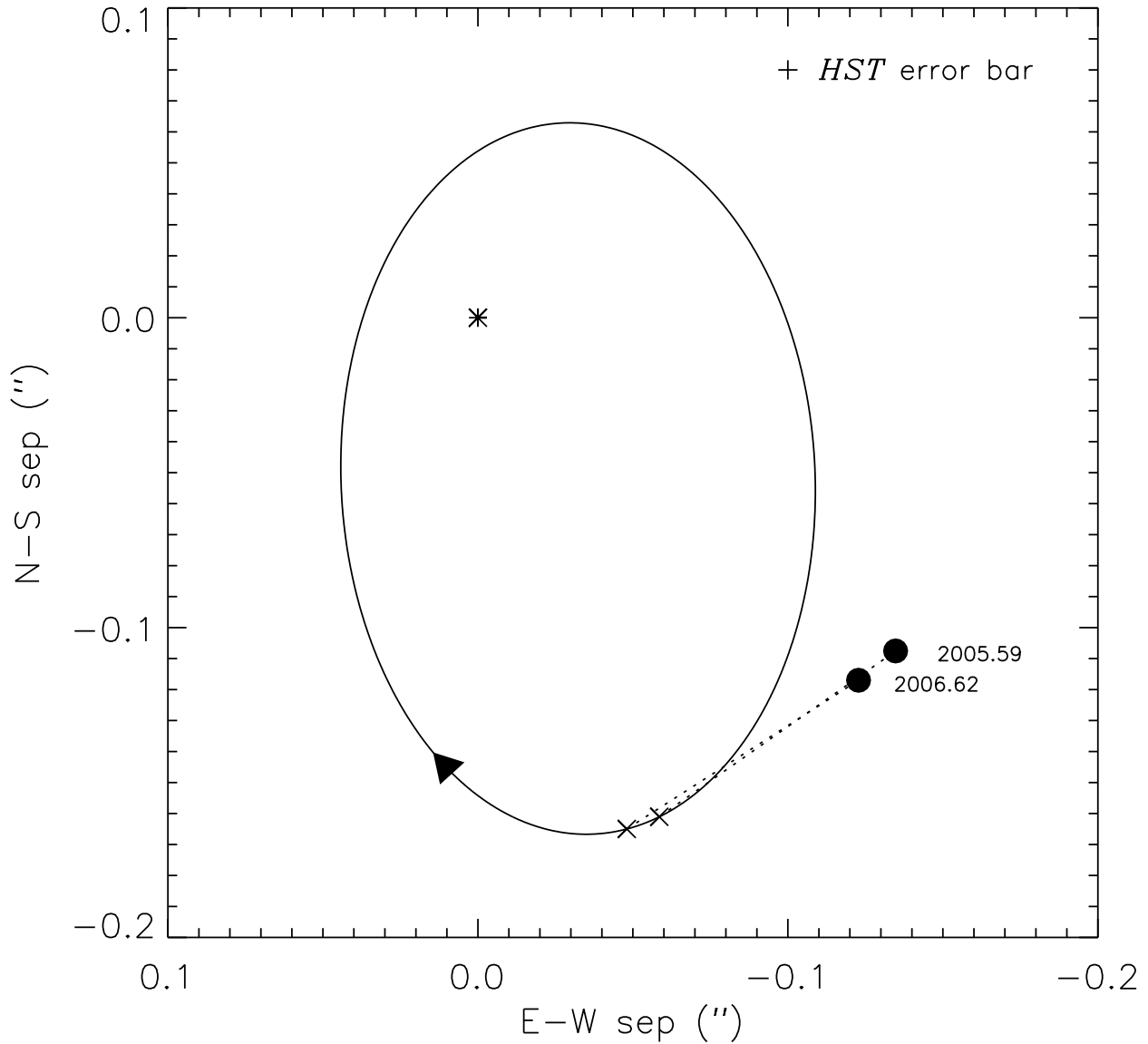}

\caption{Comparison of the retrograde orbital solution of Wielen et al.\ (2000)
(solid ellipse) with our {\it HST\/} measurements of Polaris~Ab (filled
circles).  The arrow indicates the direction of motion. The orbit was calculated
by fixing the spectroscopic and astrometric orbital parameters (see Table~3) and
solving only for the semi-major axis based on the {\it HST\/} separation
measurements.  We found $a = 0\farcs131$. The dotted lines connect the observed
positions to $\times$'s marking the predicted positions, and show the inadequacy
of this simple solution at predicting the position angles. The size of the {\it
HST\/} error bar is indicated by the cross at the top of the plot.}

\end{figure}

\begin{figure}
\plotone{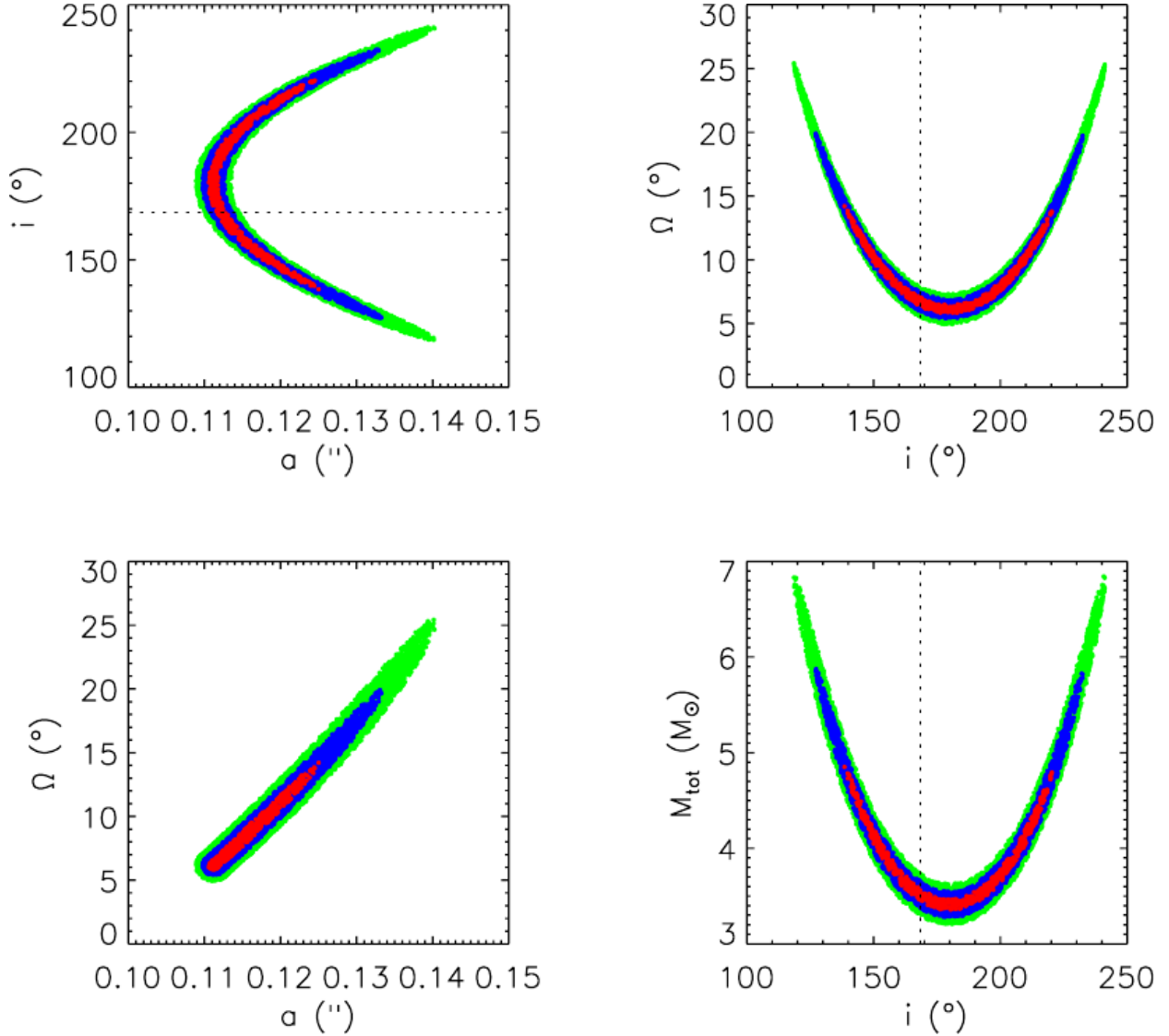}

\caption{Crosscuts through the $\chi^2$ surfaces derived from the fit to the
{\it HST\/} measurements alone while holding the spectroscopic parameters
fixed.  These results were obtained by performing a Monte-Carlo search for
orbital solutions within the $3 \sigma$ confidence interval ($\Delta\chi^2 =
9$). The total mass was derived for each of the 10,000 orbits found within this
interval using a parallax of 7.72 mas.  The dotted lines indicate the critical
value of the inclination ($i=168^\circ$), above which negative values of a
component mass are produced.  The color codes in the electronic version of this
figure correspond to the $1 \sigma$ (red), $2\sigma$ (blue), and $3 \sigma$
(green) confidence intervals.}

\end{figure}

\begin{figure}
\plotone{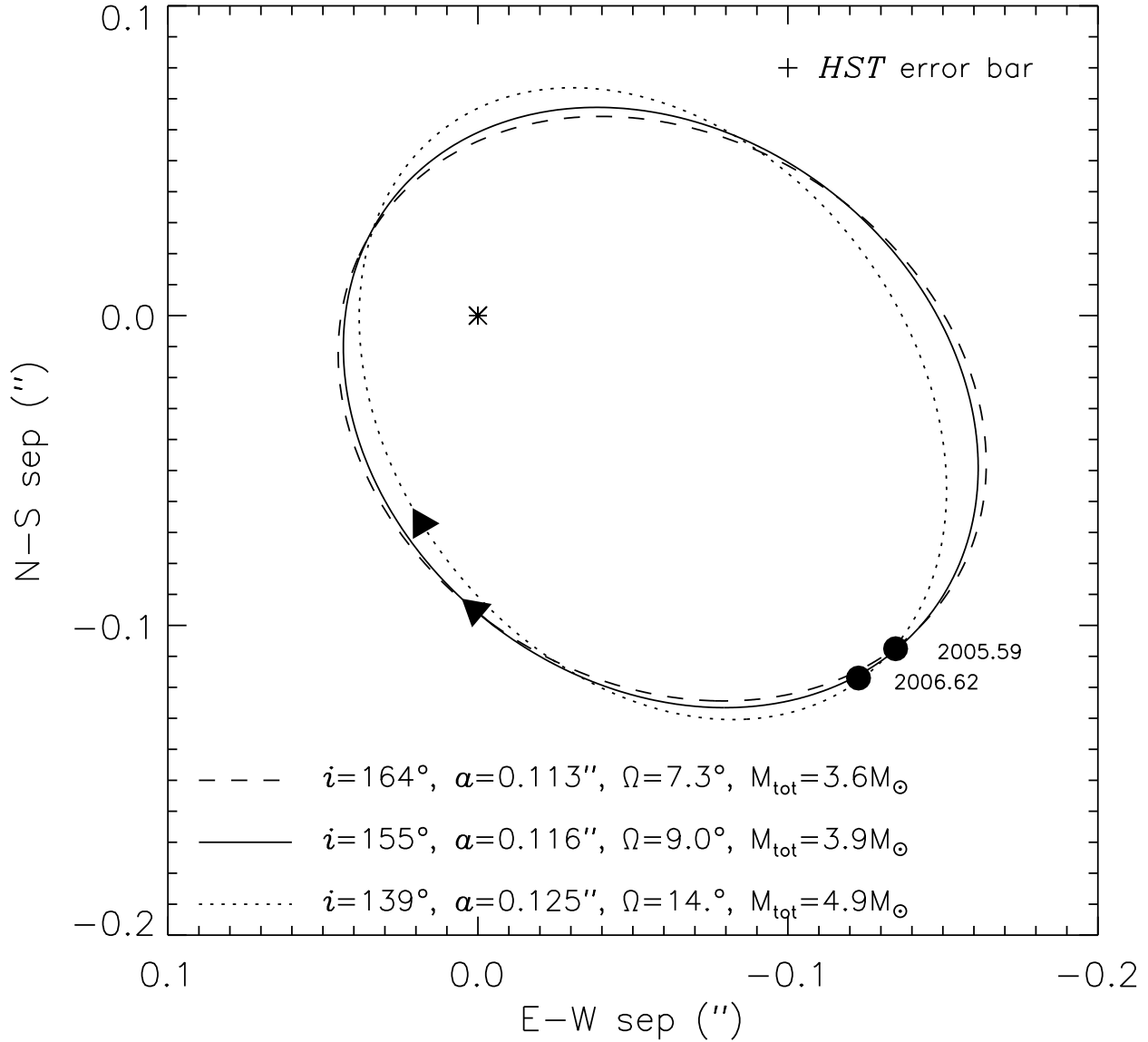}

\caption{Three examples of orbits resulting from variations of the orbital
parameters within the $1\sigma$ confidence intervals based on the fit to the
{\it HST\/} measurements alone while holding the spectroscopic parameters fixed.
The arrows indicate the direction of motion. All three orbits fit the {\it
HST\/} measurements (filled circles) within the $1\sigma$ error bar shown at the
top of the figure, but they imply total system masses ranging from 3.6 to
$4.9\,M_\odot$.}

\end{figure}

\begin{figure}
\plotone{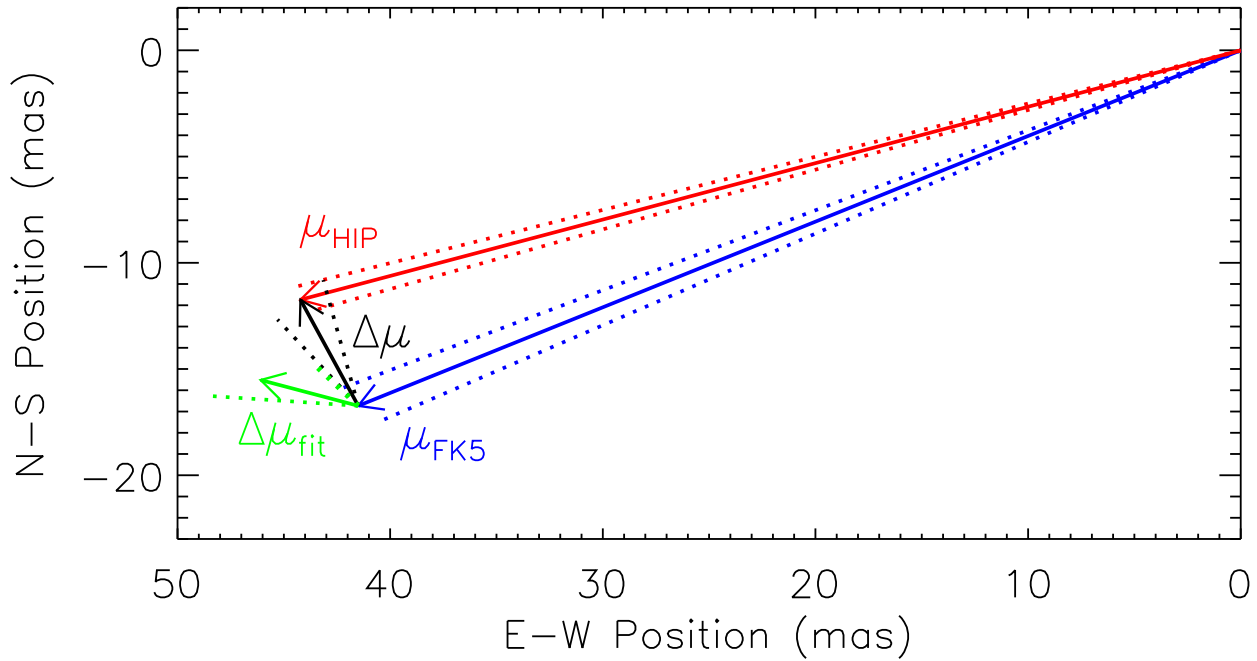}

\caption{Schematic comparison between the instantaneous and mean proper motions
of Polaris~A\null.  The vectors indicate the magnitude and direction of the
annual proper-motion measurements from {\it Hipparcos\/} ($\mu_{\rm HIP}$;
red), the long-term average FK5 ground-based motion ($\mu_{\rm FK5}$; blue), and
the vector difference between the two ($\Delta \mu$; black).  Dotted red and
blue lines indicate the $1\sigma$ uncertainties in the {\it Hipparcos\/} and FK5
measurements. The green vector represents the best-fit difference between the
proper motions  ($\Delta \mu_{\rm fit}$) computed from our simultaneous orbit
fit to the {\it HST\/} measurements and the proper-motion data while holding the
spectroscopic parameters fixed (see \S3.3).} 

\end{figure}

\begin{figure}
\plotone{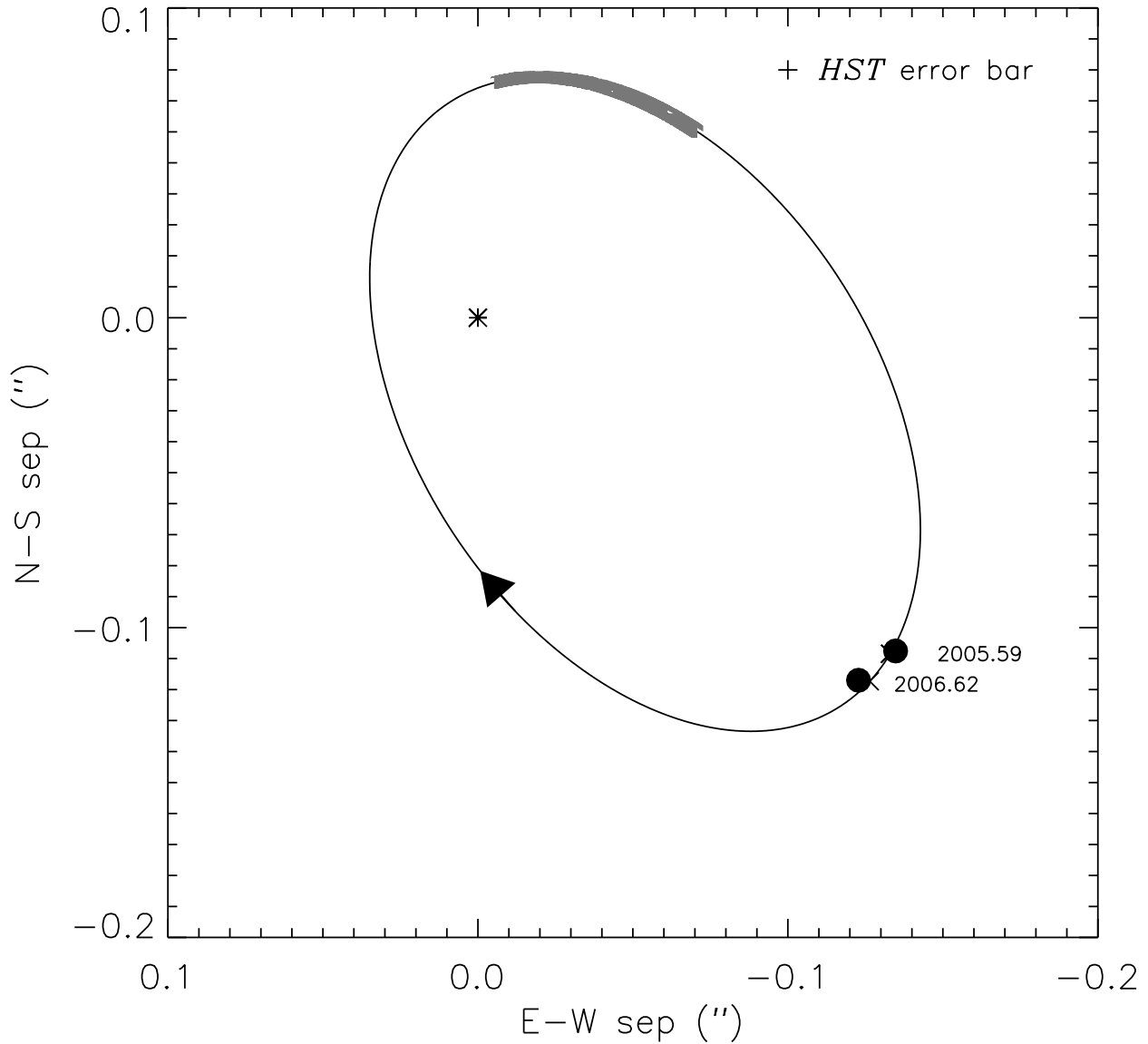}

\caption{Final best-fit orbit of Polaris Ab relative to Aa based on a 
simultaneous fit to the {\it HST\/} measurements (filled circles) and the
proper-motion data while holding the spectroscopic parameters fixed. The
predicted orbital positions at the epochs of the {\it HST\/} measurements are
marked by the $\times$ symbols (partially hidden by the observed points). The
position of the companion during the time-frame of the {\it Hipparcos\/} mission
is highlighted by the shaded gray line segment; its direction of motion is of
course $180^\circ$ different from the direction of $\Delta\mu_{\rm fit}$, shown
as a green arrow in Figure~7.}

\end{figure}

\begin{figure}
\plotone{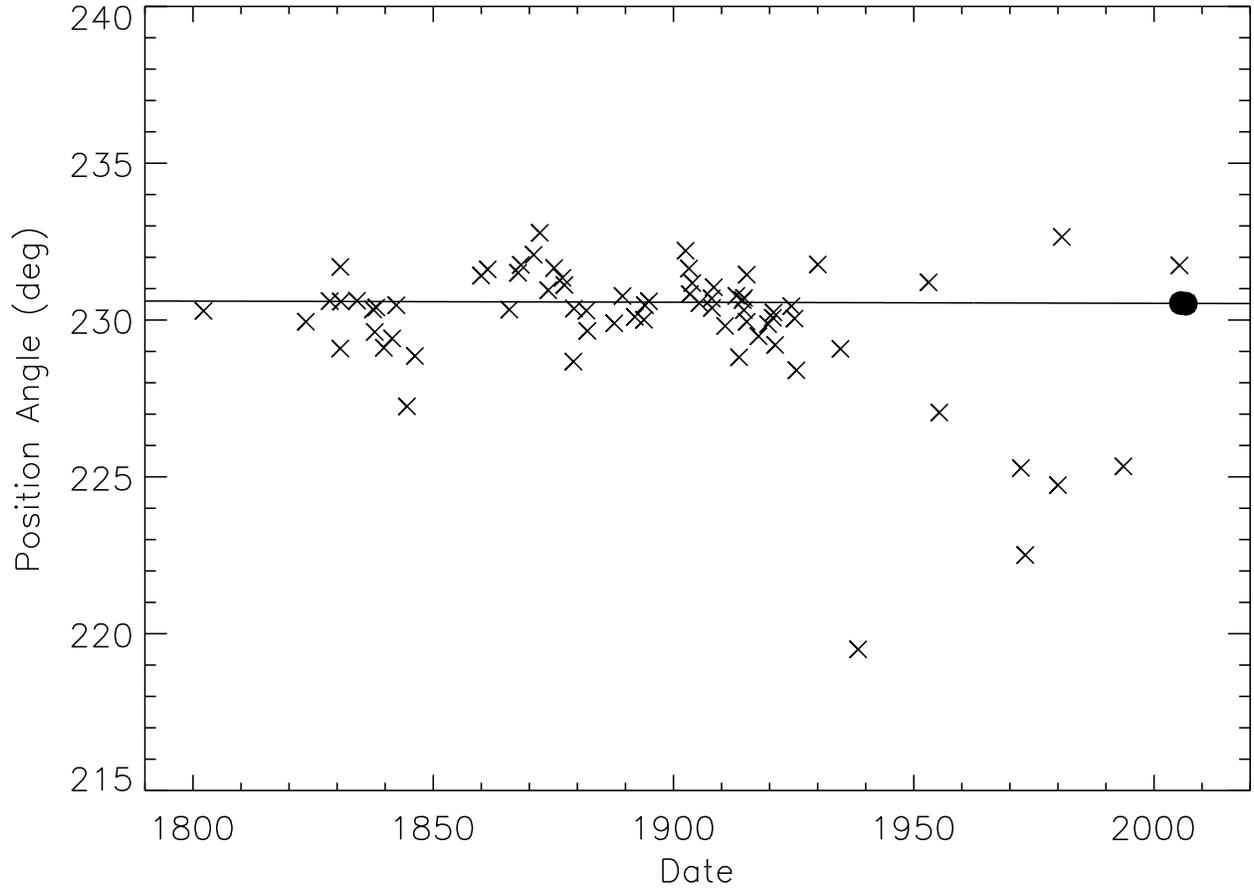}

\caption{Historical measurements of the position angle of Polaris B relative to
A, precessed to the equinox of J2000.0. Our {\it HST\/} measurements are marked
by the two filled circles. The solid line shows a least-squares fit yielding a
rate of change in the position angle of $-0\fdg00035 \pm  0\fdg00094\rm
~yr^{-1}$.}

\end{figure}

\begin{figure}
\plotone{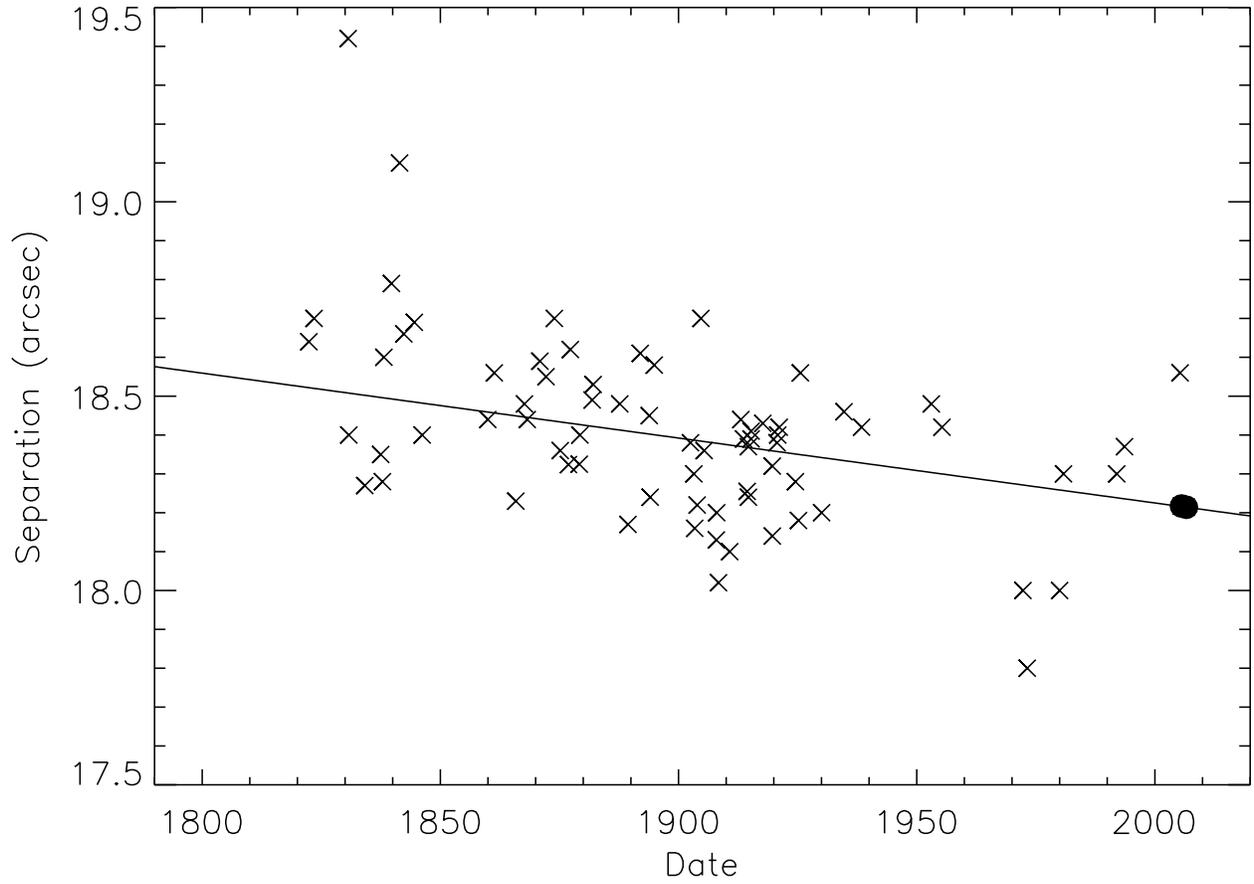}

\caption{Separation measurements of Polaris B relative to A.  Our {\it HST\/}
measurements are marked by the two filled circles.  The solid line shows a
linear least-squares fit yielding a rate of change in the separation of
$-1.67\pm0.19\rm\,mas\,yr^{-1}$.}

\end{figure}

\begin{figure}
\plotone{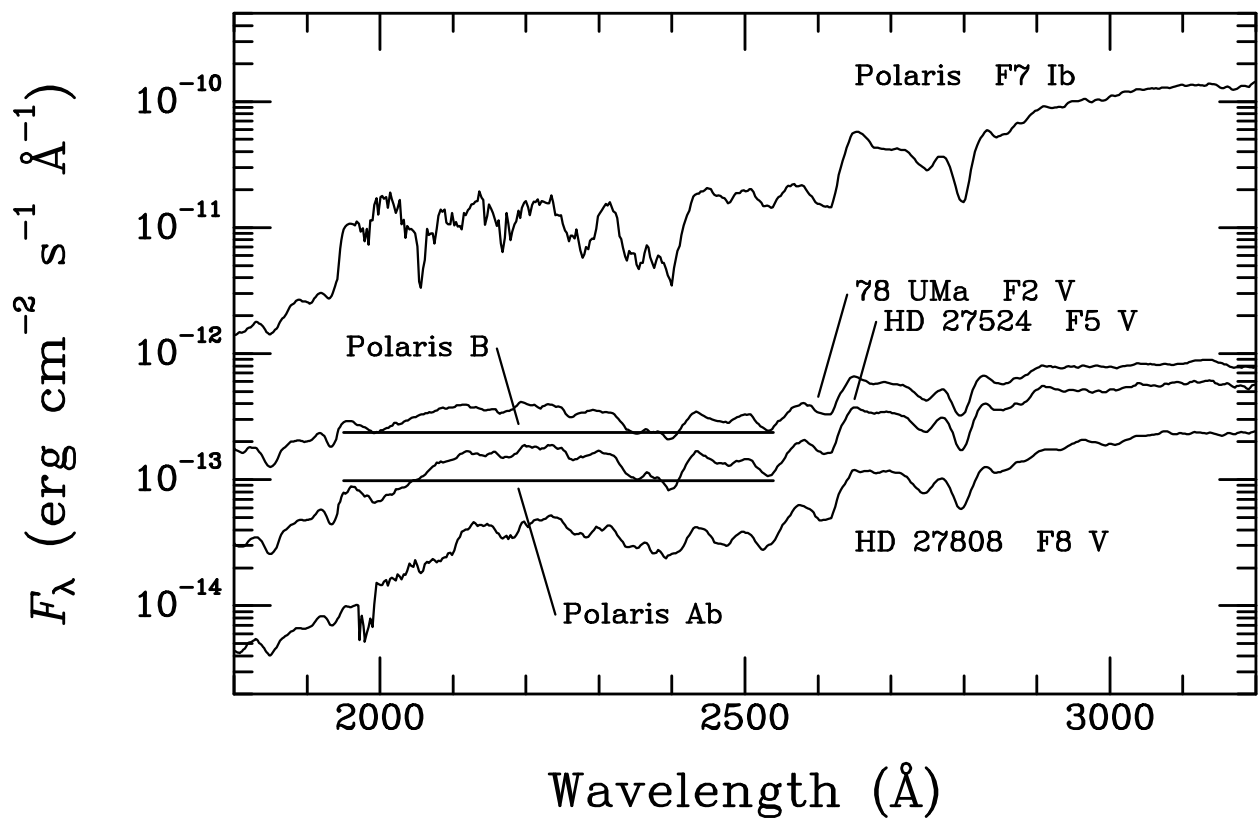}

\caption{ {\it IUE\/} UV spectra of Polaris, and of three main-sequence F-type
stars scaled to the distance of Polaris using the respective {\it Hipparcos\/}
parallaxes. The two horizontal lines correspond to the mean flux levels in the
F220W filter of the ACS/HRC of Polaris~B (top) and Ab (bottom). The lengths of
the horizontal lines correspond to the FWHM of the filter. On the basis of its
UV flux level, Polaris~B is inferred to have a spectral type near F3-F4~V, in
good agreement with its ground-based classification at F3~V\null. Polaris~Ab is
inferred from its F220W flux to have a spectral type near F6~V.}

\end{figure}

\begin{figure}  
\epsscale{0.7}
\plotone{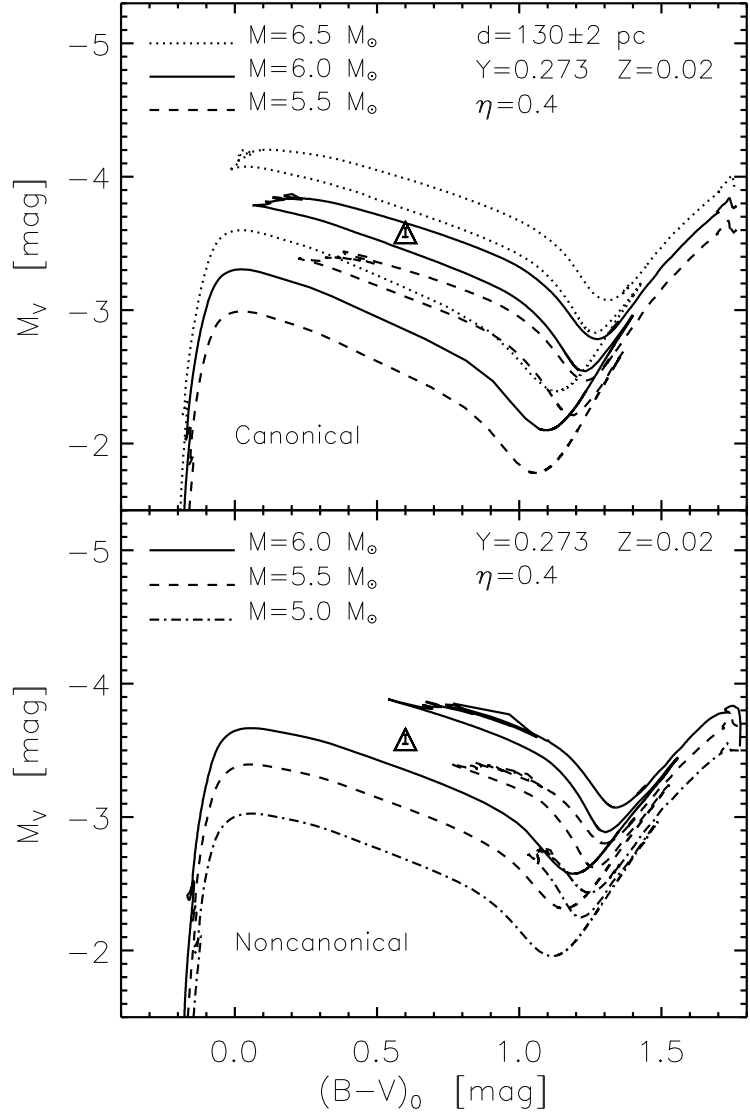}
\vskip0.5in

\caption{Evolutionary tracks (see Pietrinferni et al.\ 2006) in the $M_V$ vs.\
$(B-V)_0$ plane, with the location of Polaris marked ({\it open triangle\/}
enclosing error bar). {\it Top panel:} canonical evolutionary models  neglecting
convective core overshooting during H-burning phases. {\it Bottom panel:}
noncanonical evolutionary models including convective core overshooting. The
assumed Polaris distance and helium ($Y$) and metal ($Z$) abundances (scaled
solar) are indicated in the figure, and the legends on the left indicate the
stellar masses for each track. These models include mass loss with a Reimers
parameter $\eta=0.4$.}

\end{figure}

\clearpage

\begin{deluxetable}{llccc}  
\tablewidth{0pt}           
\tablecaption{Position and magnitude of Polaris Ab relative to Polaris Aa}  
\tablehead{\colhead{Besselian Date} & \colhead{UT Date \& Time} & 
  \colhead{$\rho$ $('')$} & \colhead{PA (J2000) ($^\circ$)\tablenotemark{a}} & 
  \colhead{$\Delta m$(F220W)} }
\startdata                                                               
2005.5880 & 2005 Aug 2  23:45 & 0.172 $\pm$ 0.002 & 231.4 $\pm$ 0.7 & 
  5.38 $\pm$ 0.09 \\  
2006.6172 & 2006 Aug 16 22:01 & 0.170 $\pm$ 0.003 & 226.4 $\pm$ 1.0 & 
  5.40 $\pm$ 0.09 \\  
\enddata
\tablenotetext{a}{PAs for equinox of date are $232\fdg8$ and $228\fdg1$.}
\end{deluxetable}

\begin{deluxetable}{llccc}     
\tablewidth{0pt}              
\tablecaption{Position and magnitude of Polaris B relative to Polaris Aa}   
\tablehead{\colhead{Besselian Date} & \colhead{UT Date \& Time} & 
  \colhead{$\rho$ $('')$} & \colhead{PA (J2000) ($^\circ$)\tablenotemark{a}} & 
  \colhead{$\Delta m$(F220W)} }
\startdata                    
2005.5880 & 2005 Aug 2 23:45  & 18.217 $\pm$ 0.003 & 230.540 $\pm$ 0.009 & 
  4.53 $\pm$ 0.04 \\    
2006.6172 & 2006 Aug 16 22:01 & 18.214 $\pm$ 0.003 & 230.520 $\pm$ 0.009 & 
  4.45 $\pm$ 0.02 \\    
\enddata                      
\tablenotetext{a}{PAs for equinox of date are $231\fdg980$ and $232\fdg216$.}
\end{deluxetable}

\begin{deluxetable}{|cc|cc|cc|}
\tablewidth{0pt}
\tablecaption{Previously determined orbital parameters}
\tablehead{&&&&&\\[-23pt]
&&&&&\\[-12pt]
\multicolumn{2}{|c|}{Kamper (1996)} & 
  \multicolumn{2}{c|}{Wielen et al.\ (2000)} & 
  \multicolumn{2}{c|}{Wielen et al.\ (2000)} \\ 
\multicolumn{2}{|c|}{ } & \multicolumn{2}{c|}{Prograde} & 
  \multicolumn{2}{c|}{Retrograde} }
\startdata
&&&&&\\[-24pt]
&&&&&\\[-12pt]
$P$ (yr) & \phn\phn29.59 $\pm$ 0.02\phn & $i~(^\circ)$ & \phn50.1 $\pm$ 4.8 & $i~(^\circ)$ & 
  130.2 $\pm$ 4.9  \\
$T$ & 1987.66 $\pm$ 0.13\phn & $\Omega$\tablenotemark{a} \ $(^\circ)$ & 
  276.2 $\pm$ 9.5 & $\Omega$\tablenotemark{a} \ $(^\circ)$ & 167.1 $\pm$ 9.4 \\
$e$ & \phn\phn0.608 $\pm$ 0.005  &  & & & \\
$\omega$ $(^\circ)$    & \phn303.01 $\pm$ 0.75\phn & & & & \\
$K_{\rm Aa}$ (km s$^{-1}$) & \phn\phn\phn3.72 $\pm$ 0.03\phn   & & & & \\[-3pt]
\enddata
\tablenotetext{a}{Values of $\Omega$ quoted in Wielen et al.\  (2000) correspond
to the astrometric orbit of Polaris Aa relative to the center of mass.}
\end{deluxetable}


\begin{deluxetable}{lccc}
\tablewidth{0pt}
\tablecaption{Orbital Parameters\tablenotemark{a} ~and Masses\tablenotemark{b}}
\tablehead{\colhead{Parameter} & \colhead{Wielen et al.\ (2000)} & 
 \colhead{Fit {\it HST}} & \colhead{Joint Fit to {\it HST}}  \\
 & \colhead{Retrograde\tablenotemark{c}} & \colhead{only} & 
  \colhead{\& Proper Motion} }
\startdata
$i~(^\circ)$                           & 130.2 (F)      & 
  $155^{+14}_{-16}$   & $128^{+14}_{-21}$     \\
$\Omega$\tablenotemark{d} $\,(^\circ)$ & 347.1 (F)      & 
  $9.0^{+5.3}_{-2.6}$ & 19$^{+15}_{-7}$       \\
$a~('')$                               & $0.131\pm0.04$ & 
  $0.116^{+0.009}_{-0.006}$ & 0.133$^{+0.015}_{-0.011}$ \\
$M_{\rm tot}\,(M_\odot)$  	       & $\phn\phn5.6\pm1.0\phn$    & 
  $3.9^{+1.0}_{-0.5}$ & $5.8^{+2.2}_{-1.3}$	\\
$M_{\rm Aa} \,(M_\odot)$  	       & $\phn\phn4.3\pm1.3\phn$    & 
  $2.1^{+1.4}_{-2.1}$ & $4.5^{+2.2}_{-1.4}$   \\
$M_{\rm Ab} \,(M_\odot)$  	       & $\phn1.26\pm0.80$  & 
  $1.8^{+1.8}_{-0.5}$ & $1.26^{+0.14}_{-0.07}$ \\
\enddata
\tablenotetext{a}{Spectroscopic orbital parameters $(P,T,e,\omega)$ held fixed
  at Kamper (1996) values.}
\tablenotetext{b}{Assumes {\it Hipparcos\/} parallax of $7.72 \pm 0.12$ mas.}
\tablenotetext{c}{Orbital parameters marked (F) are fixed at the given
  values when computing the best-fit solution.}
\tablenotetext{d}{$\Omega$ has been rotated by 180$^\circ$ from the values
  quoted in Wielen et al.\ (2000) to correspond to the orbit of Polaris Ab
  relative to Polaris Aa.}
\end{deluxetable}

\begin{deluxetable}{lccc}
\tablewidth{0pt}
\tablecaption{Proper-motion Data for Polaris A}
\tablehead{\colhead{Quantity} & \colhead{System} & 
 \colhead{$\mu_\alpha\cos{\delta}$} & \colhead{$\mu_\delta$} \\
\colhead{} & \colhead{} & \colhead{(mas yr$^{-1}$)} & 
 \colhead{(mas yr$^{-1}$)} }
\startdata
$\mu_{\rm FK5}$ & FK5 & +38.30 $\pm$ 0.23 & $-15.20 \pm 0.35$ \\
$\mu_{\rm FK5}$ & HIP & +41.50 $\pm$ 0.97 & $-16.73 \pm 0.75$ \\
$\mu_{\rm HIP}$   & HIP & +44.22 $\pm$ 0.47 & $-11.74 \pm 0.55$ \\
$\Delta\mu = \mu_{\rm HIP} - \mu_{\rm FK5}$ & HIP & \phn+2.72 $\pm$ 1.08 & 
  \phn+4.99 $\pm$ 0.93 \\
$\Delta\mu_{\rm fit}$ & HIP & \phn+4.59 $\pm$ 2.52 & \phn+1.21 $\pm$ 0.74 \\
\enddata
\end{deluxetable}

\begin{deluxetable}{lccccc}
\tablecaption{Spectral-Type Comparisons}
\tablewidth{0pt}
\tablehead{
\colhead{Star} & \colhead{Spectral} & \colhead{Mass} & \colhead{Parallax} &
  \colhead{$M_V$}  & \colhead{F220W Flux}  \\
\colhead{} & \colhead{Type} & \colhead{($M_\odot$)} & \colhead{(mas)} &
  \colhead{ } & \colhead{Relative to Polaris\tablenotemark{a}} }
\startdata
78 UMa     & F2 V    & 1.41    & 40.06   & +2.9    & 0.0262 \\
Polaris B  & $\dots$ & $\dots$ & $\dots$ & $\dots$ & 0.0178  \\
HD 27524   & F5 V    & 1.33    & 19.55   & +3.2    & 0.0114 \\
Polaris Ab & $\dots$ & $\dots$ & $\dots$ & $\dots$ & 0.0074 \\
HD 27808   & F8 V    & 1.22    & 24.47   & +4.1    & 0.0029 \\
\enddata
\tablenotetext{a}{Flux ratios for 78~UMa, HD~27524, and HD~27808 are predicted
from their {\it IUE\/} spectra, scaled to the distance of Polaris; ratios for
Polaris~B and~Ab are those observed by us, corrected for red leak as described
in the text.}
\end{deluxetable}


\begin{thebibliography}{}

\bibitem[anderson and king 2004]{ak04} Anderson, J., \& King, I. R. 2004, 
Instrument Science Report ACS 2004-15 (Baltimore: STScI)

\bibitem[]{} Bono, G., Caputo, F., Cassisi, S., Marconi, M., Piersanti, L., \&
Tornamb\`e, A. 2000, \apj, 543, 955

\bibitem[]{} Bono, G., Caputo, F., \& Castellani, V. 2006, MemSAIt, 77, 207 
 
\bibitem[]{} Bono, G., Castellani, V., \& Marconi, M. 2002, \apj, 565, L83 

\bibitem[]{} Bono, G., Gieren, W. P., Marconi, M., \& Fouque, P. 2001a, \apj,
552, L141 

\bibitem[]{} Bono, G., Gieren, W. P., Marconi, M., Fouque, P., \& Caputo, F.
2001b, \apj, 563, 319

\bibitem[]{} Bruntt, H., et al.\ 2008, \apj, in press; ArXiv e-prints, 804,
arXiv:0804.3593 

\bibitem[Caputo 2005]{Ca05} Caputo, F., Bono, G., Fiorentino, G., Marconi, M.,
\& Musella, I. 2005, \apj, 629, 1021

\bibitem[]{} Castor, J. I., Abbott, D. C., \& Klein, R. I. 1975, \apj, 195, 157

\bibitem[]{} Chiaberge, M., \& Sirianni, M. 2007, Instrument Science Report ACS
2007-03 (Baltimore: STScI)

\bibitem[]{} Chiosi, C., Bertelli, G., \& Bressan, A. 1992, ARA\&A, 30, 235 

\bibitem[]{} Evans, N.~R.\ 1988, \pasp, 100, 724 

\bibitem[]{} Evans, N. R., Barmby, P., Marengo, M., Bono, G., Welch, D.,
\& Romaniello, M. 2007, BAAS, 39, 112

\bibitem[]{} Evans, N. R., Massa, D., Fullerton, A., Sonneborn, G., \& Iping,
R. 2006, \apj, 647, 1387

\bibitem[]{} Feast, M. W., \& Catchpole, R. M. 1997, \mnras, 286, L1

\bibitem[]{} Fernie, J.~D.\ 1966, \aj, 71,  732 

\bibitem[]{} Fernie, J. D., Evans, N., Beattie, B., \& Seager, S. 1995, IBVS,
4148, 1 

\bibitem[Harmanec 1988]{ph88} Harmanec, P. 1988, Bull. Ast. Inst. 
Czech., 39, 329

\bibitem[Kamper 1996]{kk96} Kamper, K. W. 1996, JRASC, 90, 140

\bibitem[]{} Keller, S. C. 2008, ApJ, 677, 483

\bibitem[]{} Keller, S. C., \& Wood, P. R. 2006, \apj, 642, 834

\bibitem[]{} Kervella, P., M\'erand, A., Perrin, G., \& Coude Du Foresto, V.
 2006, A\&A, 448, 623 

\bibitem[]{} Kervella, P., M{\'e}rand, A., Szabados, L., Fouqu{\'e}, P.,
Bersier, D., Pompei, E., \& Perrin, G.\ 2008, \aap, 480, 167 

\bibitem[]{} Luck, R.~E., \& Bond, H.~E.\ 1986, \pasp, 98, 442 

\bibitem[]{} Mason, B. D., Wycoff, G. L., Hartkopf, W. I., Douglass, G. G., \&
Worley, C. E. 2001, AJ, 122, 3466

\bibitem[]{} McNamara, D.~H.\ 1969, \pasp,  81, 68 

\bibitem[]{} M\'erand, A., et al. 2006, A\&A, 453, 155 

\bibitem[]{} Meynet, G., \& Maeder, A. 2000, \aap, 361, 101 

\bibitem[]{} Natale, G., Marconi, M., \& Bono, G. 2008, \apj, 674, L93  

\bibitem[]{} Neilson, H. R., \& Lester, J. B. 2008, ApJ, in press;
 arXiv:0803.4198v1

\bibitem[]{} Nieuwenhuijzen, H., \& de Jager, C. 1990, A\&A, 231, 134  

\bibitem[]{} Nordgren, T. E., Armstrong, J. T., Germain, M. E., Hindsley, R.
B., Hajian, A. R., Sudol, J. J., \& Hummel, C. A. 2000, \apj, 543, 972

\bibitem[]{} Pietrinferni, A., Cassisi, S., Salaris, M., \& Castelli, F. 2006,
\apj, 642, 797

\bibitem[]{} Reimers, D.\ 1975, Memoires of the Societe Royale des Sciences de
Liege, 8, 369 

\bibitem[]{} Stothers, R. B., \& Chin, C.-W. 1991, \apj, 374, 288 

\bibitem[Turner 1977]{tu77} Turner, D. G. 1977, \pasp, 89, 550

\bibitem[]{} Turner, D.~G., Savoy, J., Derrah, J., Abdel-Sabour Abdel-Latif,
M., \& Berdnikov, L.~N.\ 2005, \pasp, 117, 207 

\bibitem[]{} Turner, D.~G., Usenko, I.~A., Miroshnichenko, A.~S., Klochkova,
V.~G., Panchuk, V.~E., \& Yang, S.~L.\ 2006, Bulletin of the American
Astronomical Society, 38, 118 

\bibitem[]{} Usenko, I., \& Klochkova, V.\ 2007, ArXiv e-prints, 708,
arXiv:0708.0333 

\bibitem[]{} Usenko, I.~A., Miroshnichenko, A.~S., Klochkova, V.~G., \&
Yushkin, M.~V.\ 2005, \mnras, 362, 1219  

\bibitem[]{} van Leeuwen, F., Feast, M.~W., Whitelock, P.~A., \& Laney, C.~D.\
2007, \mnras, 379, 723 

\bibitem[Wielen etal 2000]{wetal77} Wielen, R., Jahrei{\ss}, H., Dettbarn, C.,
Lenhardt, H., \& Schwan, H.\ 2000, \aap, 360, 399

\bibitem[]{} Xu, H. Y., \& Li, Y. 2004, \aap, 418, 225 

\end{thebibliography}
\end{document}